\documentclass[aps,pra,twocolumn,superscriptaddress]{revtex4-1}
\usepackage[latin9]{inputenc}
\usepackage{amsmath}
\usepackage{amssymb}
\usepackage{graphicx}
\usepackage[unicode=true,
 bookmarks=false,
 breaklinks=false,pdfborder={0 0 1},colorlinks=false]
 {hyperref}
\hypersetup{
 dvipdfm,colorlinks}



\begin{document}

\title{Quantum simulation of Hofstadter butterfly with synthetic gauge fields
on two-dimensional superconducting-qubit lattices}

\author{Wei Feng }

\affiliation{School of Physics, Hangzhou Normal University, Hangzhou 311121, China}

\author{Dexi Shao }

\affiliation{School of Physics, Hangzhou Normal University, Hangzhou 311121, China}

\author{Guo-Qiang Zhang }

\affiliation{School of Physics, Hangzhou Normal University, Hangzhou 311121, China}

\author{Qi-Ping Su}

\affiliation{School of Physics, Hangzhou Normal University, Hangzhou 311121, China}

\author{Jun-Xiang Zhang }
\thanks{Corresponding author: junxiang\_zhang@zju.edu.cn}

\affiliation{School of Physics, Zhejiang University, Hangzhou 310027, China}

\author{Chui-Ping Yang}
\thanks{Corresponding author: yangcp@hznu.edu.cn}

\affiliation{School of Physics, Hangzhou Normal University, Hangzhou 311121, China}

\date{\today }
\begin{abstract}
Motivated by recent realizations of two-dimensional (2D) superconducting-qubit
lattices, we propose a protocol to simulate Hofstadter butterfly with
synthetic gauge fields in superconducting circuits. Based on the existing\textbf{
}2D superconducting-qubit lattices, we construct a generalized Hofstadter
model on zigzag lattices, which has a fractal energy spectrum similar
to the original Hofstadter butterfly. By periodically modulating the
resonant frequencies of qubits, we engineer a synthetic gauge field
to mimic the generalized Hofstadter Hamiltonian. A spectroscopic method
is used to demonstrate the Hofstadter butterfly from the time evolutions
of experimental observables. We numerically simulate the dynamics of the system
with realistic parameters, and the results show a butterfly spectrum
clearly. Our proposal provides a promising way to realize the Hofstadter
butterfly on the latest 2D superconducting-qubit lattices and will
stimulate the quantum simulation of novel properties induced by magnetic
fields in superconducting circuits.
\end{abstract}
\maketitle

\section{introduction}

Quantum simulation is promising to efficiently solve many-body problems
of complex quantum systems \cite{Feynman}. In general, there are
two strategies in quantum simulation: one is digital quantum simulation
based on quantum logic gates, and the other is analog quantum simulation
which directly mimics the target Hamiltonian with controllable quantum
systems \cite{RMP2014,Buluta2009}. Compared with digital quantum
simulation, analog quantum simulation is not universal, but it has
higher tolerance level of errors \cite{Somaroo1999}. For many problems
that require only qualitative but not quantitative results, analog
quantum simulations can give meaningful results even if they are affected
by quantum decoherence. Experimental demonstrations of analog quantum
simulation have already been performed on a number of platforms, such
as optical lattices, trapped ions, and so on \cite{Simon2011,Bloch2012,kim2010}.
It is expected that special-purpose analog quantum simulators for
some practical tasks will be realized in the near future.

Hofstadter butterfly is a famous fractal energy spectrum that was
first predicted in theory by Hofstadter in 1976 \cite{Hofstadter1976}.
The original Hofstadter butterfly was found in a system of electrons
on a two-dimensional (2D) square lattice exposed to a perpendicular
magnetic field. In general, the experimental observation requires
that the characteristic length of the magnetic field is comparable
to the lattice constant. The estimated strength of the magnetic field
required to observe the Hofstadter butterfly for conventional solid-state
crystals is on the order of $10^{4}\thinspace\mathrm{T}$, which is
too high to be experimentally realized. One way to overcome this problem
is to use graphene superlattices with larger lattice constant, which
can reduce the required strength of magnetic fields \cite{Dean2013,Ponomarenko2013,Hunt2013,Wu2021,Rozhkov2016}.
On the other hand, the Hofstadter butterfly can be demonstrated by
analog quantum simulation with synthetic gauge fields in artificial
systems. When a charged particle moves on a closed loop under a magnetic
field, its wave function will obtain a geometric phase related to
the magnetic flux through the loop. Synthesizing a gauge field can
be equated to trying to make a neutral particle acquire a non-trivial
phase after one revolution around a closed loop. The synthesis of
gauge fields in artificial quantum systems is of great importance
to simulate magnetic effects including the Hofstadter butterfly. In
recent years, the implementation of synthetic gauge fields in artificial
controllable systems (e.g., cold atoms \cite{Dalibard2011,Galitaki2013,Gerbier2010},
cavity QEDs \cite{Cho2008,Carusotto2011}, superconducting circuits
\cite{Roushan2017,Girvin2010,Girvin2011,Zoller2013,HuY2016,Hu2016,Alaeian2019,Zhao2020,Guan2020})
and related quantum simulations have become a hot research area. Simultaneously,
many protocols have been proposed to mimic the Hofstadter model with
synthetic gauge fields in platforms such as optical lattices with
ultracold atoms \cite{Zoller2003}, trapped ions \cite{Grab2015},
and polaritonic systems \cite{Banerjee2018}. In experiments \cite{Aidelsburger2013,Miyake2013},
researchers provided evidence that they have engineered the Hofstadter
Hamiltonian, but the fractal energy spectrum could not be measured
directly.

Superconducting circuits provide an excellent platform to perform
quantum simulations due to their flexibility and scalability \cite{CQED-Review,Gu2017,CQED-review2,Malley2016,Xu2018,Feng2022,Su2022,LiuTong2022,Su2021,LiuTong2020}.
Recently, significant technological breakthroughs have been made in
superconducting circuits. In particular, the experimental realization
of 2D architectures, such as square \cite{lattice1,lattice2,lattice3,lattice4,lattice5},
zigzag \cite{zigzag1,zigzag2,zigzag3}, ladder \cite{zhu2019}, and
hyperbolic \cite{Kollar2019} lattices, makes it possible to simulate
2D lattice models using superconducting circuits. For the simulation
of Hofstadter butterfly on 2D superconducting-qubit lattices, a qubit
flip interaction is analog to the hopping of an electron. The key
problem is how to create the synthetic gauge fields. One method for
engineering the synthetic gauge field in superconducting circuits
is based on the periodic modulation of interaction strengths \cite{Roushan2017,Girvin2010,Girvin2011,Zoller2013,HuY2016,Hu2016},
also named as parametric process, which however can not be applied
on the recent 2D superconducting-qubit lattices \cite{lattice1,lattice2,lattice3,lattice4,lattice5,zigzag1,zigzag2,zigzag3,zhu2019}.
By contrast, another method is based on the periodic modulation of
on-site energies, which is easy to be realized for superconducting
transmon qubits and has been adopted in the protocols of generating
synthetic gauge fields on ladder lattices \cite{Alaeian2019,Zhao2020,Guan2020}.
Although a reduced 1D Harper Hamiltonian \cite{Harper1955,Das2019}
has been simulated using a chain of superconducting qubits \cite{Martinis2017science},
the simulation of 2D Hofstadter Hamiltonian in superconducting circuits
is still lacking.

In this paper, motivated by experimental realizations of 2D lattices
in superconducting qubit circuits \cite{lattice1,lattice2,lattice3,lattice4,lattice5,zigzag1,zigzag2,zigzag3,zhu2019,Kollar2019},
we theoretical study the Hofstadter butterfly with synthetic gauge fields on 2D
superconducting-qubit lattices. It is difficult to realize a uniform
gauge field on the existing square lattices of superconducting qubits.
Therefore, we generalize the original Hofstadter model from square
lattices to zigzag lattices, and we find a fractal energy spectrum
similar to the original Hofstadter butterfly. To mimic the generalized
Hofstadter Hamiltonian, we design a modulation scheme based on the
zigzag superconducting-qubit lattices previously reported. By modulating
the qubit frequencies with appropriate initial phases and amplitudes,
we engineer an artificial magnetic field and realize the Hofstadter
Hamiltonian on zigzag superconducting-qubit lattices. Further, we
resolve the fractal energy spectrum from time evolutions of experimental
observables. We also numerically simulate the evolution process based
on the original time-dependent Hamiltonian of the superconducting-qubit
lattice. With a group of realistic parameters, a simulated spectrum
resembles a butterfly obviously, which confirms our scheme is feasible
with existing superconducting-qubit lattices.

This paper is organized as follows: In Sec. \ref{sec:model}, we introduce
the Hofstadter model on zigzag lattices and show its fractal energy
spectrum. In Sec. \ref{sec:synthetic-magnetic-field}, we present
the modulation scheme for generating artificial gauge fields and mimicking
the Hofstadter Hamiltonian on zigzag superconducting-qubit lattices.
In Sec. \ref{sec:observing-fractal-energy}, we give a method to detect
the fractal energy spectrum from time evolutions of experimental observables
and discuss the experimental feasibility with numerical simulations.
Finally, we make a conclusion in Sec. \ref{sec:conclusion}.

\section{model \label{sec:model}}

\begin{figure}
\begin{centering}
\includegraphics[width=8cm]{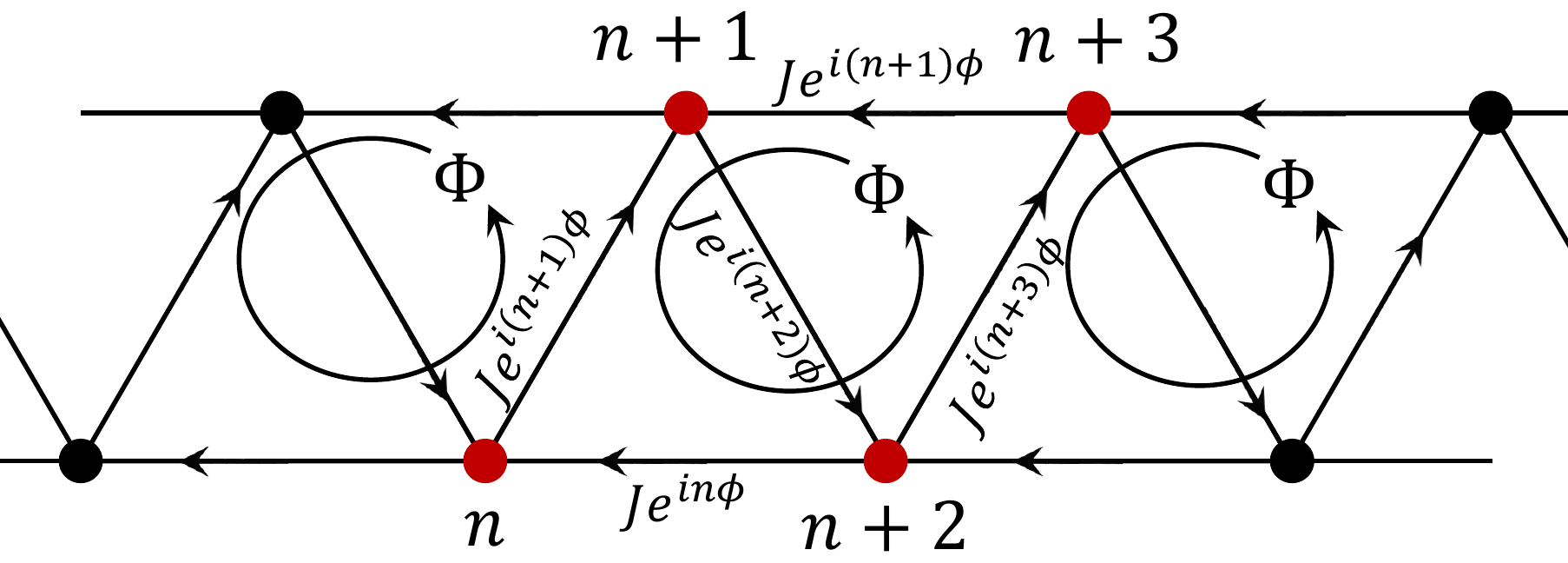}
\par\end{centering}

\caption{A zigzag spin chain with nearest-neighbor interactions. Arrows between
spins denote the complex-valued interactions shown in Eq. (\ref{eq:1}).
The shown setup leads to constant fluxes $\Phi=3\phi$ through rhombic
unit cells. \label{fig:1} }
\end{figure}

\begin{figure}[t]
\begin{centering}
\includegraphics[width=8.5cm]{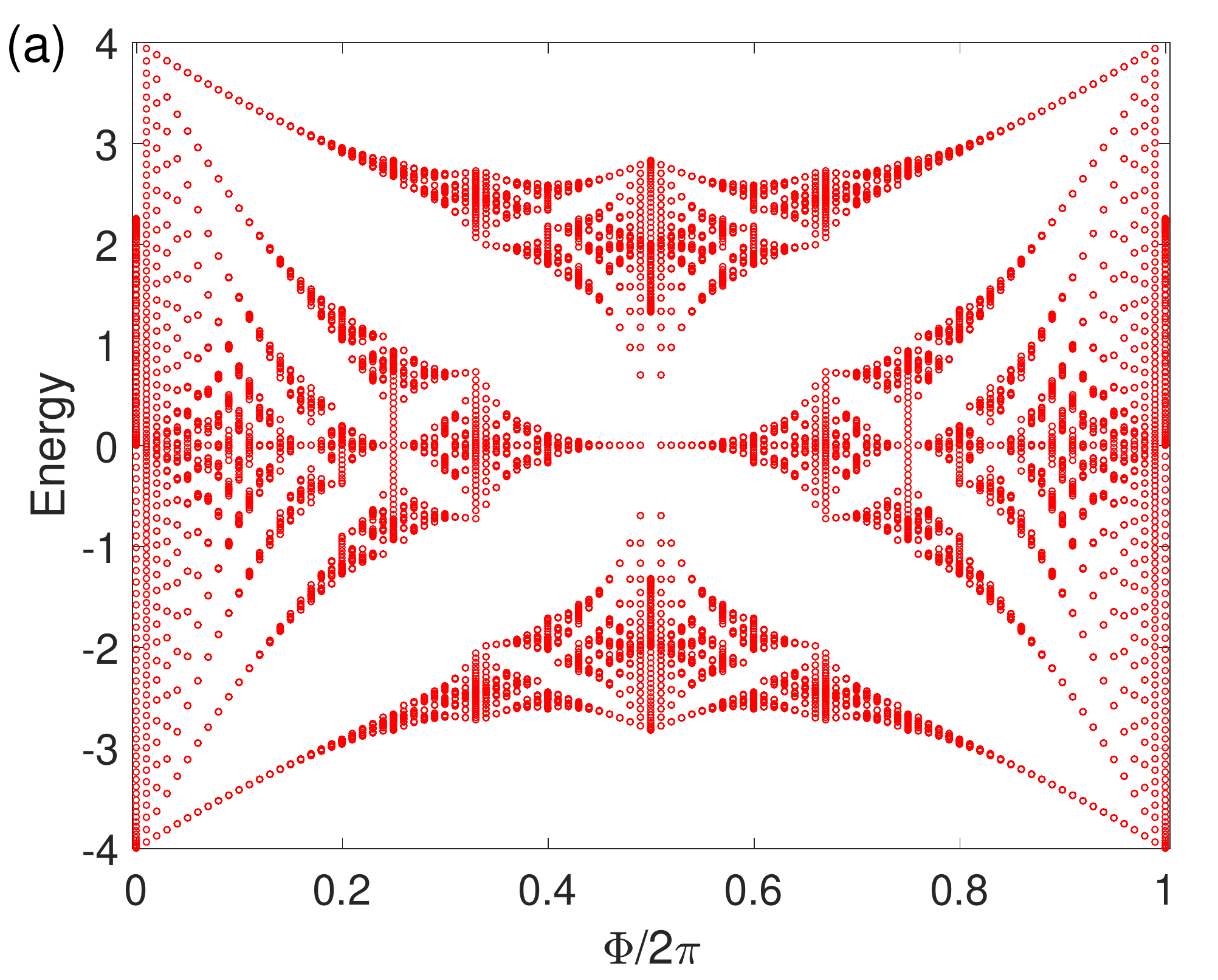}
\par\end{centering}

\begin{centering}
\includegraphics[width=8.5cm]{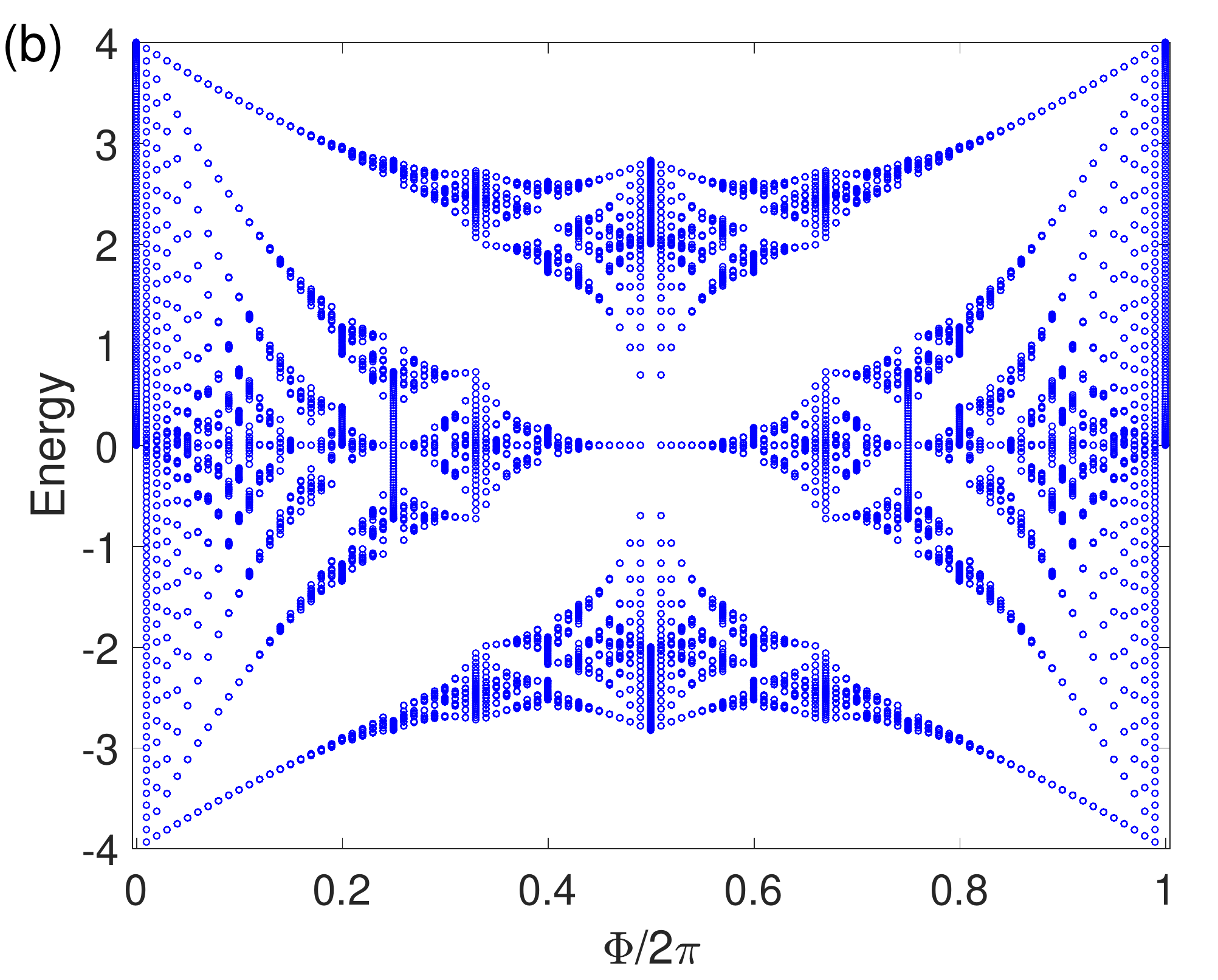}
\par\end{centering}

\caption{Fractal energy spectra: the energy eigenvalues (in units of $J$)
are plotted as a function of the flux $\Phi$ (in units of $2\pi$)
in the single spin-flip subspace. (a) The fractal energy spectrum
of the generalized model on the zigzag lattice. (b) The original Hofstadter
butterfly on the square lattice. We adopt periodic boundary conditions
in a system of $N=300$ spins. \label{fig:2}}
\end{figure}

We consider a zigzag spin lattice, which is experimentally achievable
in superconducting circuits \cite{zigzag1,zigzag2,zigzag3}. As shown
in Fig. \ref{fig:1}, each spin is coupled to four nearest neighbors
and the coupling coefficients are complex numbers with spatially varying
phases. The system can be described by a Hofstadter-like Hamiltonian
(assuming $\hbar=1$)

\begin{equation}
H=J\sum_{n}\left[e^{in\phi}\sigma_{n}^{+}\sigma_{n+2}^{-}+e^{-i\left(n+1\right)\phi}\sigma_{n}^{+}\sigma_{n+1}^{-}\right]+\mathrm{H.c.},\label{eq:1}
\end{equation}
where $\sigma_{n}^{+}$($\sigma_{n}^{-}$) is the raising (lowering)
operator of the $n$th spin, and $J$ represents the coupling strength
between spins. This Hamiltonian commutes with the total excitation
$\sum_{n}\sigma_{n}^{+}\sigma_{n}^{-}$ and the system is energy-conserving.
In the single-excitation manifold, i.e., only one spin is up and all
others are down, this spin-flip model maps onto a free hopping model
of a charged particle in a lattice under magnetic fields. For a triangle
formed by three spins $n$, $n+1$, and $n+2$, the associated magnetic
flux $\left(3n+3\right)\phi$ is spatially dependent. However, the
flux through rhombic unit cells, as shown in Fig. \ref{fig:1}, is
a constant: $\Phi=3\phi$. When $\Phi/2\pi$ is rational, i.e., $\Phi/2\pi=p/q$
with $p$ and $q$ two coprime integers, the Hamiltonian is of period
$q.$ In the momentum space, the dispersion $E\left(k\right)$ has
$q$ bands with the wave vector $k$ restricted to a magnetic Brillouin
zone $\left[-\pi/q,\pi/q\right]$. By Fourier transform $\sigma_{n}^{+}=\sum_{k,v}e^{-in\left(k+v\phi\right)}\sigma_{k,v}^{+}$,
with $v$ the band index, the Hamiltonian is transformed to $H=\sum_{k}H_{k}$,
where
\begin{equation}
H_{k}=J\sum_{v}\left[e^{i\left(k+v\phi\right)}+e^{-i2\left(k+v\phi\right)}\right]\sigma_{k,v}^{+}\sigma_{k,v+1}^{-}+\mathrm{H.c.}.
\end{equation}
In contrast to the cosine modulation of the on-site energies in the
original Harper Hamiltonian \cite{Harper1955,Das2019,Martinis2017science}
\begin{equation}
H_{\mathrm{Harper}}=J\sum_{n}\sigma_{n}^{+}\sigma_{n+1}^{-}+\sigma_{n+1}^{+}\sigma_{n}^{-}+2\cos\left(n\Phi\right)\sigma_{n}^{+}\sigma_{n}^{-},\label{eq:Harper}
\end{equation}
the coupling strengths between the neighbouring bands in $H_{k}$
periodically vary with the magnetic flux $\Phi$ as
\begin{equation}
\left|e^{i\left(k+v\phi\right)}+e^{-i2\left(k+v\phi\right)}\right|=\sqrt{2+2\cos\left(3k+v\Phi\right)}.
\end{equation}
We solve the model by diagonalizing a $N\times N$ matrix with $N$
the number of the spins. The eigenenergy spectrum versus the magnetic
flux is plotted in Fig. \ref{fig:2}(a). For comparison, an original
Hofstadter butterfly obtained by diagonalizing $H_{\mathrm{Harper}}$
in Eq. (\ref{eq:Harper}) is plotted in Fig. \ref{fig:2}(b). One
can see that the fractal energy spectrum of the zigzag lattice is
very similar to the original Hofstadter butterfly, despite small difference
at, e.g., $\Phi=0$. The small difference of the energy spectra can
be clearly explained through the Hamiltonian matrices of $H_{k}$
and $H_{\mathrm{Harper}}$. When $\Phi=0$, the matrix trace of $H_{\mathrm{Harper}}$
is $2N$ (in units of $J$) while the matrix trace of $H_{k}$ is zero,
which is confirmed by eigenenergy data shown in Fig. \ref{fig:2}.

\section{synthetic gauge field\label{sec:synthetic-magnetic-field}}

\begin{figure}
\begin{centering}
\includegraphics[width=8.5cm]{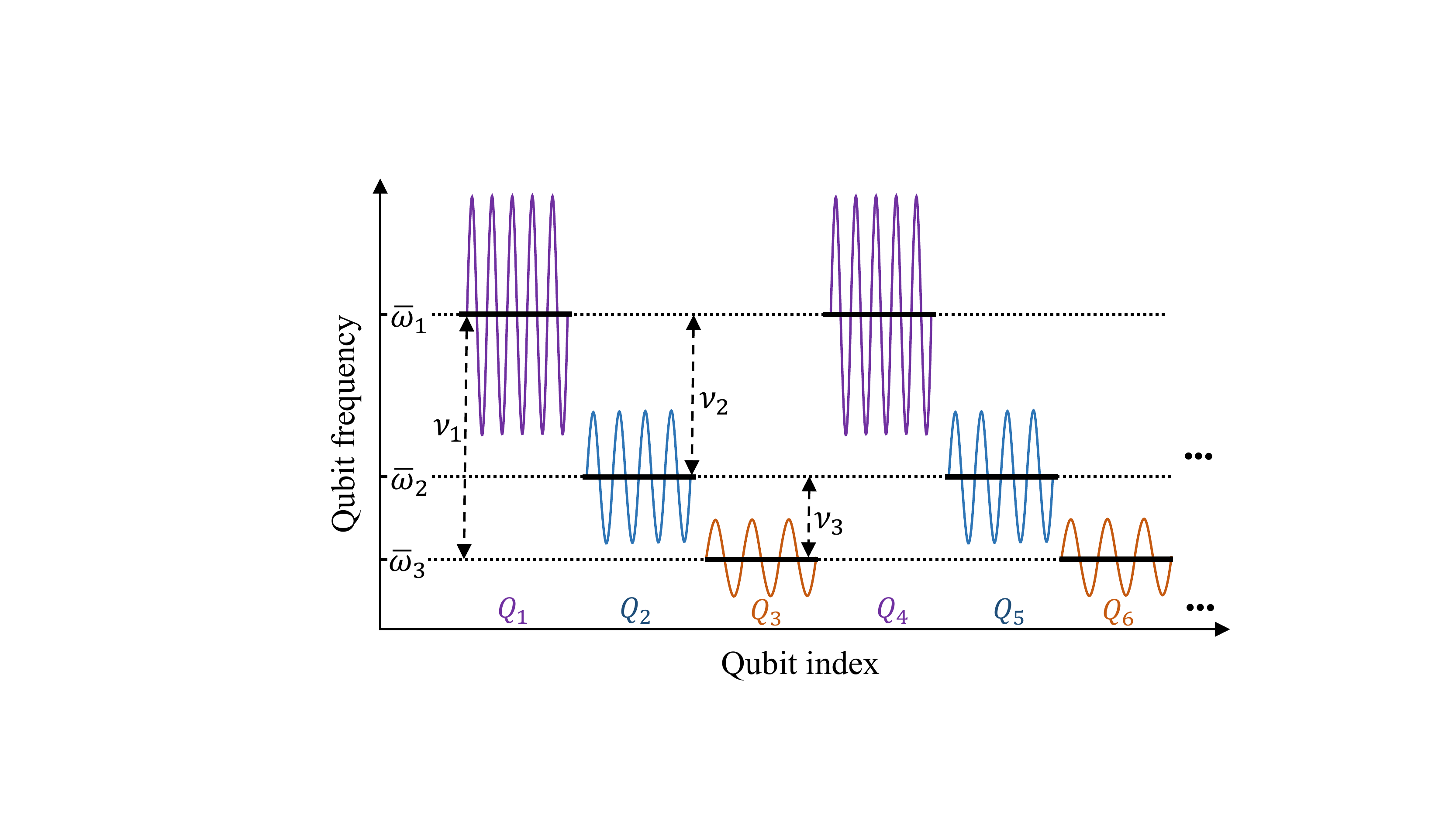}
\par\end{centering}

\caption{Frequency setting and modulation scheme for synthesizing the complex-valued
interaction. The resonant frequency of each qubit is periodically
modulated around its central frequency $\bar{\omega}_{n}$ (black
solid line). For sideband transitions, the modulation frequency is
specially setted with $\nu_{1}=\bar{\omega}_{1}-\bar{\omega}_{3}$,
$\nu_{2}=\bar{\omega}_{1}-\bar{\omega}_{2}$, and $\nu_{3}=\bar{\omega}_{2}-\bar{\omega}_{3}$.
Latter qubits repeat the settings $\bar{\omega}_{n}=\bar{\omega}_{j}$
and $\nu_{n}=\nu_{j}$ where $n\equiv j$ (mod $3$) with $j=1,2,3$.
\label{fig:scheme}}
\end{figure}

The coupling strength between superconducting qubits is usually a
real number due to time-reversal symmetry. We can realize the effective
complex-valued coupling in Eq. (\ref{eq:1}) by modulating resonant
frequencies of qubits. The original Hamiltonian of $N$ frequency-tunable
qubits arranged in a zigzag lattice is \cite{zigzag1,zigzag2,zigzag3}

\begin{equation}
H_{o}=\sum_{n}\omega_{n}\left(t\right)\sigma_{n}^{+}\sigma_{n}^{-}+g\sum_{n}\left(\sigma_{n+1}^{+}\sigma_{n}^{-}+\sigma_{n+2}^{+}\sigma_{n}^{-}\right)+\mathrm{H.c.},\label{eq:oH}
\end{equation}
where $\sigma_{n}^{+}$($\sigma_{n}^{-}$) denotes the raising (lowering)
operator of the $n$th qubit $Q_{n}$, and $g$ is the original real-valued
coupling strength between qubits. The resonant frequency of $Q_{n}$
is periodically modulated according to
\begin{equation}
\omega_{n}\left(t\right)=\bar{\omega}_{n}+\varepsilon_{n}\cos\left(\nu_{n}t+\theta_{n}\right),\label{eq:modulate}
\end{equation}
where $\bar{\omega}_{n}$ is the central frequency, and $\varepsilon_{n}$,
$\nu_{n}$, and $\theta_{n}$ are the modulation amplitude, frequency,
and initial phase, respectively. Under the modulation, each qubit
generate a bunch of equally spaced sidebands with a spacing equalling
to the modulation frequency $\nu_{n}$ \cite{Blais2012,Blais2013,sunluyan2018statetransfer}.
We set $\nu_{1}=\Delta_{13}$, $\nu_{2}=\Delta_{12}$, $\nu_{3}=\Delta_{23}$
and repeat the settings $\bar{\omega}_{n}=\bar{\omega}_{j}$ and $\nu_{n}=\nu_{j}$
for $n\equiv j$ (mod $3$) with $j=1,2,3$ (see Fig. \ref{fig:scheme}),
where $\Delta_{mn}=\bar{\omega}_{m}-\bar{\omega}_{n}$. In these settings,
sideband transitions with tunable qubit-qubit interactions can be
implemented. Under the condition of large detuning $\left|\Delta_{mn}\right|\gg g$,
high-frequency oscillating terms can be neglected and the effective
Hamiltonian in the interaction picture can be written as \cite{sunluyan2018statetransfer}

\begin{equation}
H_{\mathrm{eff}}=\sum_{n}J_{n,n+1}\sigma_{n+1}^{+}\sigma_{n}^{-}+J_{n+2,n}\sigma_{n}^{+}\sigma_{n+2}^{-}+\mathrm{H.c.},\label{eq:Heff}
\end{equation}
where the effective coupling strengths $J_{n,n+1}$ and $J_{n+2,n}$
are given by
\begin{equation}
J_{n,n+1}=\left\{ \begin{array}{c}
g\mathcal{J}_{0}\left(\alpha_{n}\right)\mathcal{J}_{1}\left(\alpha_{n+1}\right)e^{i\theta_{n+1}},\thinspace\text{n\ensuremath{\equiv}1,2\thinspace(\ensuremath{\mathrm{mod}}\thinspace3), }\\
g\mathcal{J}_{0}\left(\alpha_{n}\right)\mathcal{J}_{-1}\left(\alpha_{n+1}\right)e^{-i\theta_{n+1}},\thinspace n\equiv3\thinspace(\mathrm{mod}\thinspace3),
\end{array}\right.
\end{equation}

\noindent \begin{flushleft}
\begin{equation}
J_{n+2,n}=\left\{ \begin{array}{cc}
g\mathcal{J}_{0}\left(\alpha_{n+2}\right)\mathcal{J}_{-1}\left(\alpha_{n}\right)e^{-i\theta_{n}}, & n\equiv1\thinspace(\mathrm{mod}\thinspace3),\\
g\mathcal{J}_{0}\left(\alpha_{n+2}\right)\mathcal{J}_{1}\left(\alpha_{n}\right)e^{i\theta_{n}}, & n\equiv2,3\thinspace(\mathrm{mod}\thinspace3),
\end{array}\right.
\end{equation}
with $\mathcal{J}_{m}\left(\alpha_{n}\right)$ being the $m$th-order
Bessel function of the first kind. Both magnitudes and phases of the
above effective coupling strengths can be conveniently tuned by changing
$\alpha_{n}=\varepsilon_{n}/\nu_{n}$ and $\theta_{n}$. If we set
\begin{equation}
\theta_{n}=\left\{ \begin{array}{cc}
\pi-n\phi, & n\equiv1\thinspace(\mathrm{mod}\thinspace3),\\
n\phi, & n\equiv2,3\thinspace(\mathrm{mod}\thinspace3),
\end{array}\right.
\end{equation}
and all $\alpha_{n}$ to be the same value, the effective Hamiltonian
$H_{\mathrm{eff}}$ becomes the Hofstadter-like Hamiltonian in Eq.
(\ref{eq:1}).
\par\end{flushleft}

In summary, based on the original Hamiltonian of the superconducting-qubit
system in Eq. (\ref{eq:oH}), we periodically modulate the resonant
frequency of each qubit according to Eq. (\ref{eq:modulate}). By
finely setting the modulation parameters $\bar{\omega}_{n}$, $\varepsilon_{n}$,
$\nu_{n}$, and $\theta_{n}$, we approximatively obtain the Hofstadter-like
Hamiltonian in Eq. (\ref{eq:1}) under the large-detuning condition
$\left|\Delta_{mn}\right|\gg g$. The parameter setting shown above
is specially designed for synthesizing the uniform gauge field in
the Hofstadter-like Hamiltonian. Other kinds of synthetic gauge fields
can be engineered by changing the parameter setting.

\section{observing fractal energy spectrum \label{sec:observing-fractal-energy}}

\begin{figure}
\begin{centering}
\includegraphics[width=8.5cm]{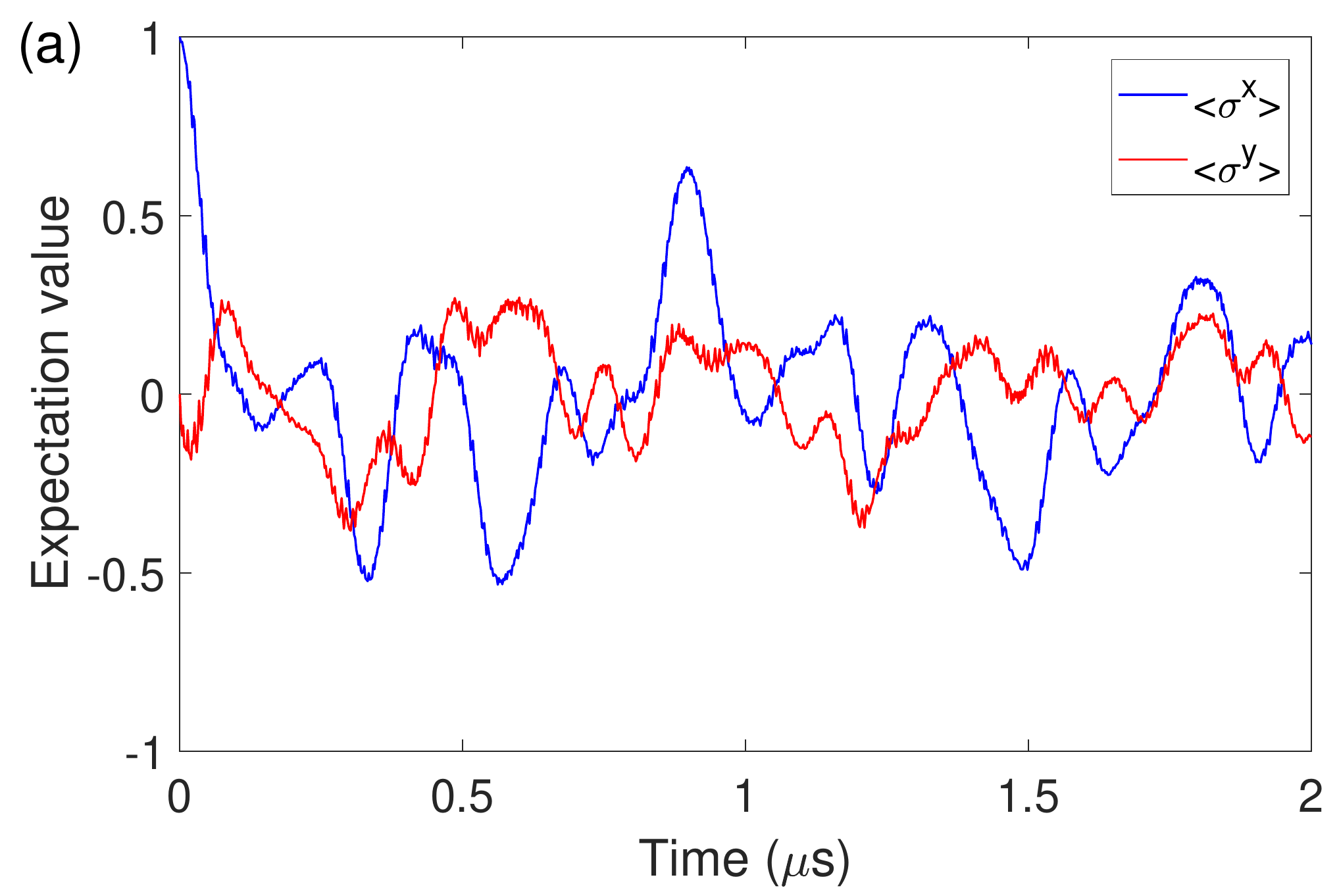}
\par\end{centering}

\begin{centering}
\includegraphics[width=8.5cm]{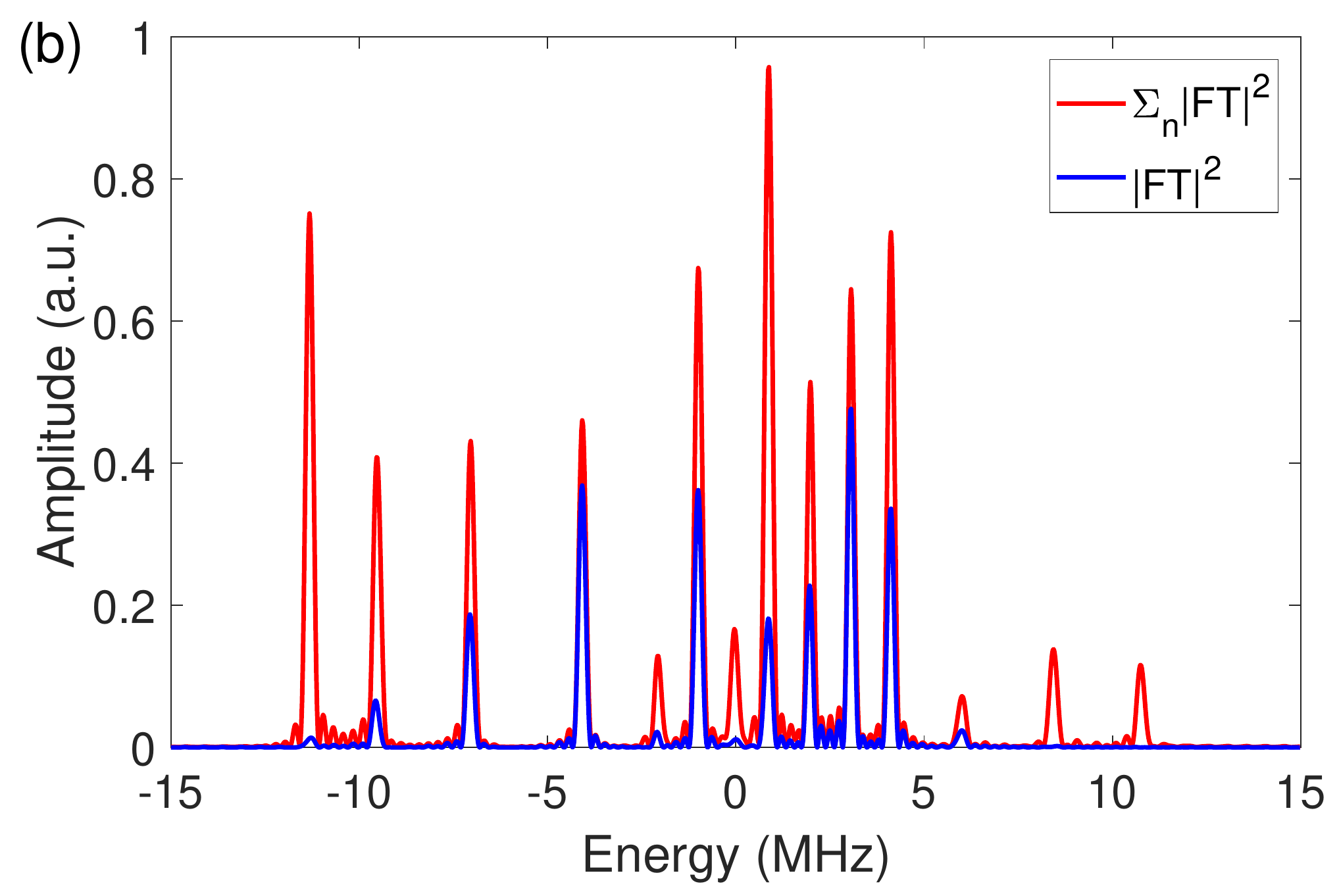}
\par\end{centering}

\caption{Spectroscopic signatures of the interacting multi-qubit system. (a)
Typical curves of time evolutions of $\left\langle \sigma_{n}^{x}\right\rangle $
and $\left\langle \sigma_{n}^{y}\right\rangle $. The results are
simulated using the original Hamiltonian $H_{o}$ with the initial
state $\left(\left|\psi_{0}\right\rangle +\left|\psi_{n}\right\rangle \right)/\sqrt{2}$
and the qubit number $N=14$. Relevant parameters are chosen as $g/2\pi=10\thinspace\mathrm{MHz}$,
$\nu_{1}/2\pi=250\thinspace\mathrm{MHz}$, $\nu_{2}/2\pi=150\thinspace\mathrm{MHz}$,
$\nu_{3}/2\pi=100\thinspace\mathrm{MHz}$, $\phi/2\pi=1/120$, and
$\alpha_{n}=1$. The relaxation and pure dephasing times of the superconducting
qubits are $T_{1}=20\thinspace\mathrm{\mu s}$ and $T_{2}^{\ast}=2\thinspace\mathrm{\mu s}$,
respectively. (b) Squared Fourier transform (FT) amplitudes of $\left\langle \sigma_{n}^{x}\right\rangle +i\left\langle \sigma_{n}^{y}\right\rangle =2\left\langle \sigma_{n}^{-}\right\rangle $.
The peaks in the FT results correspond to the eigenenergies of the
system. The blue curve is the FT result of one set of data shown in
(a), which only shows part of the eigenenergies. The red curve is
the summation of the FT results of $\left\langle \sigma_{n}^{-}\right\rangle $
for $n\in\left\{ 1,2,...,14\right\} $, which shows all $14$ eigenenergies
with $14$ peaks. \label{fig:Spectrum.}}
\end{figure}

\begin{figure*}
\begin{centering}
\includegraphics[width=8.5cm]{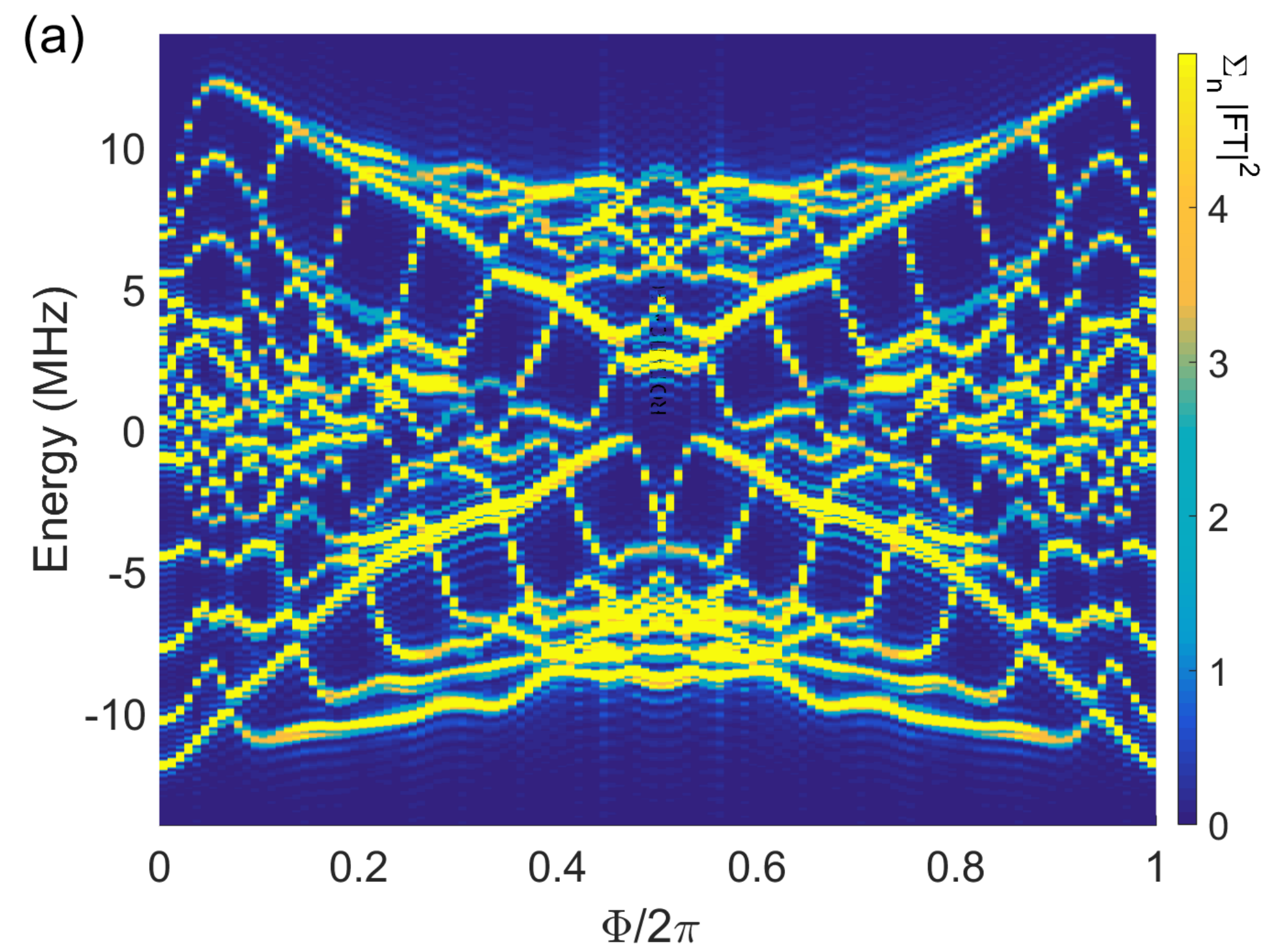} \includegraphics[width=8.5cm]{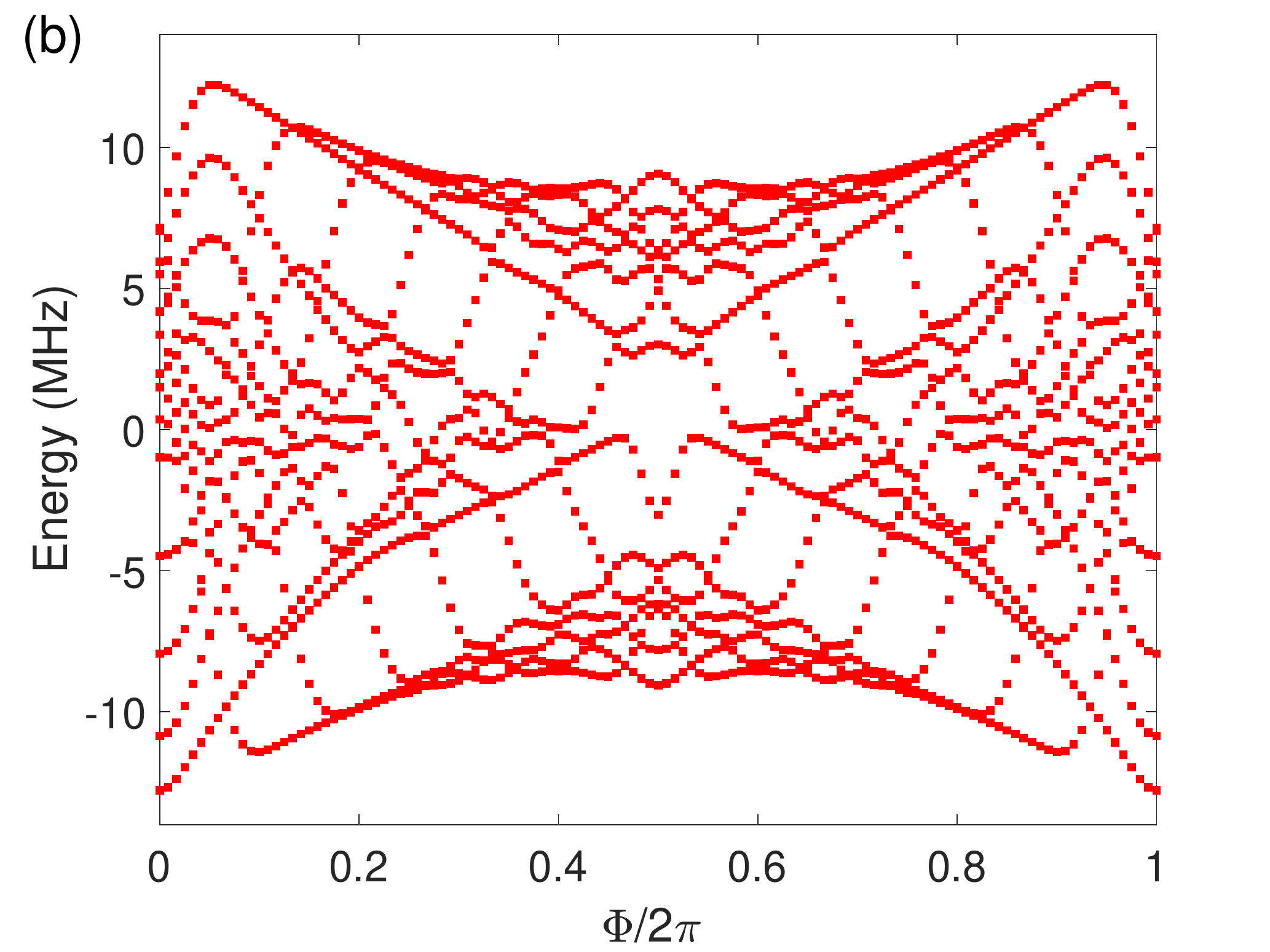}
\par\end{centering}

\caption{(a) Quantum simulation of Hofstadter butterfly on the superconducting-qubit
lattice. Data similar to Fig. \ref{fig:Spectrum.}(b), summation of
squared Fourier transform amplitudes are shown for $120$ values of
magnetic flux $\Phi/2\pi$ ranging from $0$ to $1$. For each $\Phi/2\pi$
value, time evolution of $\left\langle \sigma_{n}^{-}\right\rangle $
is simulated using the original Hamiltonian $H_{o}$ with the same
parameters as in Fig. \ref{fig:Spectrum.}(a). (b) As a comparison,
theoretical Hofstadter butterfly from model Hamiltonian $H$ is shown
with the same conditions that $N=14$ and open boundary. \label{fig:Hofstadter-Butterfly}}
\end{figure*}

The energy spectrum of an interacting many-body system can be revealed
by probing the system's dynamical responses to perturbations \cite{Martinis2017science,Spectroscopy2014,Spectroscopy2015}.
If the eigenenergies $\left\{ E_{j}\right\} $ and the corresponding
eigenstates $\left\{ \left|\psi_{j}\right\rangle \right\} $ of a
fixed Hamiltonian are given, the state time evolution of the system
can be expressed as
\begin{equation}
\left|\Psi\left(t\right)\right\rangle =\sum_{j}c_{j}e^{-iE_{j}t}\left|\psi_{j}\right\rangle ,
\end{equation}
where $c_{j}=\left\langle \psi_{j}\right|\left.\Psi\left(0\right)\right\rangle $,
and $\left|\Psi\left(0\right)\right\rangle $ is the initial state
of the system. In turn, if the time evolution is known, its Fourier
transform can in principle reveal the eigenenergies. To realize this
scheme in reality, we need to choose a group of appropriate initial
states and observables. For an operator $\hat{q}=\sum_{j,j^{\prime}}q_{j,j^{\prime}}\left|\psi_{j}\right\rangle \left\langle \psi_{j^{\prime}}\right|$,
its expectation value is
\begin{equation}
\left\langle \hat{q}\right\rangle =\left\langle \Psi\left(t\right)\right|\hat{q}\left|\Psi\left(t\right)\right\rangle =\sum_{j,j^{\prime}}q_{j,j^{\prime}}c_{j}c_{j^{\prime}}^{\ast}e^{-i\left(E_{j}-E_{j^{\prime}}\right)t},
\end{equation}
which indicates that the Fourier transforms of observable results
can only reveal eigenenergy differences $E_{j}-E_{j^{\prime}}$. In
our qubit model, the Hofstadter Butterfly is presented in the single-excitation
subspace $\left\{ \left|\psi_{n}\right\rangle \equiv\left|000\ldots1_{n}\ldots00\right\rangle \right\} $,
and the associated single-excitation eigenenergies $E_{j}$ for $j=1,2,\ldots,N$
can be calibrated by selecting the ground state ($\left|\psi_{0}\right\rangle \equiv\left|0\right\rangle ^{\otimes N}$)
energy $E_{0}\equiv0$ as a reference. To do that, the state of the
qubit system should have overlaps ($c_{j},\thinspace c_{j^{\prime}}^{\ast}$)
both with the ground state $\left|\psi_{0}\right\rangle $ and single-excitation
states $\left|\psi_{j}\right\rangle $, and the selected operator
should have corresponding matrix element $q_{j,0}$. Therefore, a
suitable initial state should be in superposition of $\left|\psi_{0}\right\rangle $
and $\left|\psi_{j}\right\rangle $, and a suitable operator should
relate to $\left|\psi_{j}\right\rangle \left\langle \psi_{0}\right|$.

Next, we verify the above scheme by numerical simulations based on
the original time-dependent Hamiltonian $H_{o}$ given in Eq. (\ref{eq:oH}).
Initially, we prepare the $n$th qubit in the state $\left(\left|0\right\rangle +\left|1\right\rangle \right)/\sqrt{2}$
and all other qubits in the state $\left|0\right\rangle $, i.e.,
the system is in the state $\left(\left|\psi_{0}\right\rangle +\left|\psi_{n}\right\rangle \right)/\sqrt{2}$.
We simulate the time evolution of the state of the system under the
periodical modulation described in Sec. \ref{sec:synthetic-magnetic-field}
by considering the qubit dissipation. We also calculate the expectation
value of the operator $\sigma_{n}^{-}$. Although $\sigma_{n}^{-}$
is non-Hermitian and cannot be measured directly, it can be inferred
from observables $\sigma_{n}^{x}$ and $\sigma_{n}^{y}$ as $\left\langle \sigma_{n}^{x}\right\rangle +i\left\langle \sigma_{n}^{y}\right\rangle =2\left\langle \sigma_{n}^{-}\right\rangle $.
Typical time evolutions of $\left\langle \sigma_{n}^{x}\right\rangle $
and $\left\langle \sigma_{n}^{y}\right\rangle $ in the simulation
are shown in Fig. \ref{fig:Spectrum.}(a), and the result of Fourier
transform of the corresponding expectation value $\left\langle \sigma_{n}^{-}\right\rangle $
is shown in Fig. \ref{fig:Spectrum.}(b). For the simulated system
containing $N$ qubits, there should be $N$ single-excitation eigenenergies.
From the blue curve in Fig. \ref{fig:Spectrum.}(b), we can see part
of eigenenergies in the time-domain spectrum. To resolve all the eigenenergies,
we vary $n$ from $1$ to $N$, which makes the initial states form
a complete basis. Then every eigenstate certainly has some overlap
with one of the initial states and hence its corresponding eigenenergy
can be detected.

To obtain the fractal spectrum versus the magnetic flux, we vary $\Phi/2\pi$
from $0$ to $1$ with $120$ different values. For each $\Phi/2\pi$
value, we place the $n$th qubit in the state $\left(\left|0\right\rangle +\left|1\right\rangle \right)/\sqrt{2}$,
simulate the evolution of $\left\langle \sigma_{n}^{-}\right\rangle $,
and vary $n$ from $1$ to $N$. Using the spectroscopy shown above,
we obtain a fractal energy spectrum of the Hofstadter butterfly shown
in Fig. \ref{fig:Hofstadter-Butterfly}(a), where the qubit number
is $N=14.$ Despite there are only 14 sites in the simulated lattice,
the overall energy spectrum still has a butterfly-like appearance.
To verify the accuracy of the energy spectrum based on Fourier transforms,
we diagonalize the $14\times14$ matrix of the Hamiltonian in Eq.
(\ref{eq:1}), and show the theoretical results in Fig. \ref{fig:Hofstadter-Butterfly}(b).
The similarity between these two energy spectra indicates that our scheme
for simulating Hofstadter butterfly with synthetic gauge fields is
effective. Note that there is small difference between Figs. \ref{fig:Hofstadter-Butterfly}(a)
and \ref{fig:Hofstadter-Butterfly}(b), especially in the smallest
eigenenergy when $\Phi=0$. The tiny deviations of the Fourier transform
spectrum from the theoretical results arise mainly from the large-detuning
approximation in obtaining the effective Hamiltonian in Eq. (\ref{eq:Heff}).

The above numerical simulations, which use the original time-dependent
Hamiltonian and consider the qubit dissipation, are performed based
on QuTiP \cite{qutip,qutip2}. Because the Lindblad-master-equation
simulation of the $14$-qubit system is beyond our computer's memory
capability, we use a Monte Carlo approach which is based on the state
vector instead of the density matrix. In the simulations, we adopt
the experimental achievable parameters of zigzag superconducting-qubit
lattices \cite{zigzag1,zigzag2,zigzag3}. In addition, the similar
frequency modulations have already been used in the experiments \cite{zigzag1,sunluyan2018statetransfer,Feng2019}.
Therefore, our scheme is experimentally realizable with the existing
superconducting-qubit lattices.

\section{conclusion\label{sec:conclusion}}

In summary, we propose an experimentally feasible scheme to demonstrate
fractal energy spectra on 2D superconducting-qubit lattices with synthetic
gauge fields. A generalized Hofstadter model on zigzag lattices is
studied by employing the superconducting-qubit lattices previously
realized in experiments. The model exhibits a fractal energy spectrum
similar to the original Hofstadter butterfly. We design a modulation
scheme of qubit frequencies to generate synthetic gauge fields and
mimic the generalized Hofstadter Hamiltonian on the zigzag superconducting-qubit
lattices. We present a method to detect the fractal energy spectrum
from time evolutions of experimental observables. With experimental
feasible parameters, our simulation results clearly demonstrate a
Hofstadter butterfly. The proposal provides a promising way to study
Hofstadter problems on the latest 2D superconducting-qubit lattices.
Our work will stimulate the quantum simulation of novel properties
induced by magnetic fields in superconducting circuits.
\begin{acknowledgments}
This work is supported by the National Natural Science Foundation
of China (Grants No. 12204139, No. U20A2076, No. 12204138, No. 12205069,
No. 11774076, and No. U21A20436) and the Key-Area Research and Development
Program of GuangDong province (Grant No. 2018B030326001).\end{acknowledgments}

\end{document}